\documentclass[floats,prb,aps,twocolumn]{revtex4}
\usepackage{graphicx}

\begin{document}

\title{Light and Electromagnetic Waves Teaching in Engineering Education}
\author{Roman Ya. Kezerashvili}
\address{Physics Department, New York City College of Technology, \\
The City University of New York\\
300 Jay Street, Brooklyn, NY 10201}

\begin{abstract}
Suggestion of physics laboratory exercises and discussion of the physics
laboratory curricula for engineering majors where the properties of light
and electromagnetic waves are studied in parallel. It is shown that one of
the important educational advantages of an experimental study of the
properties of microwaves as an example of electromagnetic waves
simultaneously with the properties of light are, on one hand, visualization
of the properties of microwaves, and on the other hand, provide evidences
that light is an electromagnetic wave.
\end{abstract}

\address{ }
\maketitle




\section{Introduction}

\qquad Engineering is the application of mathematics and science to develop
useful products or technologies. In other words, engineering is turning
ideas into reality. Physics is the study of the physical world and physics
is an indispensable component in engineering curricula because technology is
based on our knowledge of physical laws. Physics remains the leader of the
modern natural sciences, the theoretical basis of modern engineering and as
no any other science, promotes the development of creative and critical
thinking in future engineers. Good training in physics also provides a solid
base for lifelong learning.

\qquad Research in education in different countries shows that students at
the college and even university levels continue to hold fundamental
misunderstandings of the world around them. Science learning remains within
the classroom context and just a small percentage of students are able to
use the knowledge gained at school for solving various problems of the
larger physical world \cite{1,2}.

\qquad In most of the courses students hear lectures without strong
connections to their everyday experiences. Students usually do not have the
opportunity to form their own ideas, they rarely get a chance to work in a
way where they are engaged in discovery and building and testing models to
explain the world around them, like the scientists do.

\qquad In the last four years I started to include some `simple' conceptual
questions in the exams of the physics courses I taught. The results were at
first quite surprising: Most students performed very poorly on the
conceptual questions which most physics would consider as to be almost `too
easy', while they sometimes solved `difficult' multiple-step quantitative
problems better. Some of the `top' students with high scores on the
quantitative problems had very low scores on the conceptual part \cite{3}.
One of the questions which I asked is \textquotedblleft What are the
similarities and differences between electromagnetic waves and light? The
best short answer I had was \textquotedblleft Light is an electromagnetic
wave\textquotedblright\ without any deeper explanation of the properties and
phenomena. I had a lot of speculations about light but only a few students
mentioned that light is the part of the electromagnetic spectrum the human
eye can detect and they listed the main properties of electromagnetic waves.
This motivated me to revise physics laboratory curricula and develop physics
laboratory exercises where the properties of light and electromagnetic waves
are studied in parallel. Indeed, one of the places where active and
collaborative learning can be realized is the physics laboratory, where
students become active participants of the learning process \cite{4,5,6,7}.

When Maxwell showed that electric and magnetic fields can propagate through
space according to the classical wave equation and found the equation for
the speed of propagation of electromagnetic field to be

\begin{equation}
c=\frac{1}{\sqrt{\varepsilon _{0}\mu _{0}}},
\end{equation}

where $\varepsilon _{0}$ and $\mu _{0}$\ are electric permittivity and
magnetic permeability of free space, he evaluated the numerical value for
the speed of these electromagnetic waves by substituting the numerical
values for $\varepsilon _{0}$ and $\mu _{0}$\ and obtained the remarkable
result: m/s \cite{8}. Maxwell recognized that this result is very close to
the experimentally measured speed of light and made a great conceptual
conclusion that light is electromagnetic waves. Of course, this conclusion
does not surprise physicists today. Equation (1) is one of the great
equations in physics stemming as it does from Maxwell's electromagnetic
theory. This equation unifies three seemingly disparate fields of physics:
electricity, magnetism and optics.

\qquad Since electromagnetic wave concepts are usually unfamiliar, abstract,
and difficult to visualize, conceptual analogies from familiar light
phenomena are invaluable for teaching. Such analogies emphasize the
understanding of continuity of electromagnetic waves and support the spiral
development of student understanding. We found that the approach of teaching
the topics of electromagnetic waves and optics in parallel in the physics
laboratory is very helpful and useful.

\qquad The following experiments can be easily designed and they provide a
methodical introduction to electromagnetic theory using microwave radiation
and light: the study of the inverse square law of the dependence of the
intensity of radiation (microwave and light) on the distance, the law of
reflection and refraction, investigation of the phenomenon of polarization
and how a polarizer can be used to alter the polarization of microwave
radiation and light, studying interference by performing the double-slit
experiment for microwave radiation and light. Finally students measure the
wavelength of laser light and microwave radiation using the corresponding
versions of the Michelson's interferometer, and recognize that these two
forms of radiations differ only by the wavelength or frequency.

\qquad To perform the above mentioned experiments we are using a regular
light source or He-Ne laser for optics, and microwaves transmitter and
receiver for microwave electromagnetic radiation experiments. Today for
experiments with microwaves PASCO \cite{9}, as well as DAEDALON \cite{10}
provide excellent sets. In our laboratory we are using PASCO equipment. We
are using the same design for light and microwave experiments to demonstrate
the similarity of measurements and the only difference is that in the case
of light experiments students can see the phenomena and measure its physical
properties and in the case of the microwaves they can observe the same
phenomena through the meters reading of the intensity of the microwave
radiation.

\qquad The purpose of the present paper is to introduce laboratory curricula
for the study of the properties of light and microwaves in parallel in the
general physics laboratory course for engineering majors, especially for
electrical engineering and telecommunication students. The article is
organized as follows: Sec. II introduces the experiments for the inverse
square law of the dependence of the intensity of light and microwave
radiation on the distance from a source of electromagnetic waves, in Sec.
III we analyze and discuss the reflection and refraction experiments for
microwaves and light, the double-slit interference and polarization
experiments for light and microwaves are discussed in Sec. IV and Sec. V,
respectively. Interferometer measurements for wavelength of light and
microwaves are presented in Sec. VI, and conclusions follow in Sec. VII.

\section{Inverse Square Law for Light and Microwave Radiation}

\qquad The intensity received from the pointlike source of an
electromagnetic wave is inversely proportional to the square of the distance
from the wave source and an inverse square law can be written in the form

\begin{equation}
I=\frac{L}{4\pi r^{2}},
\end{equation}

where $L$ is the luminosity of the source. To observe this phenomenon we
designed the experiment where the photoelectric photometer has been used to
measure the intensity of light in increments of distances from the light
source. 
\begin{figure}[tbp]
\centering
\includegraphics[width = 85mm]{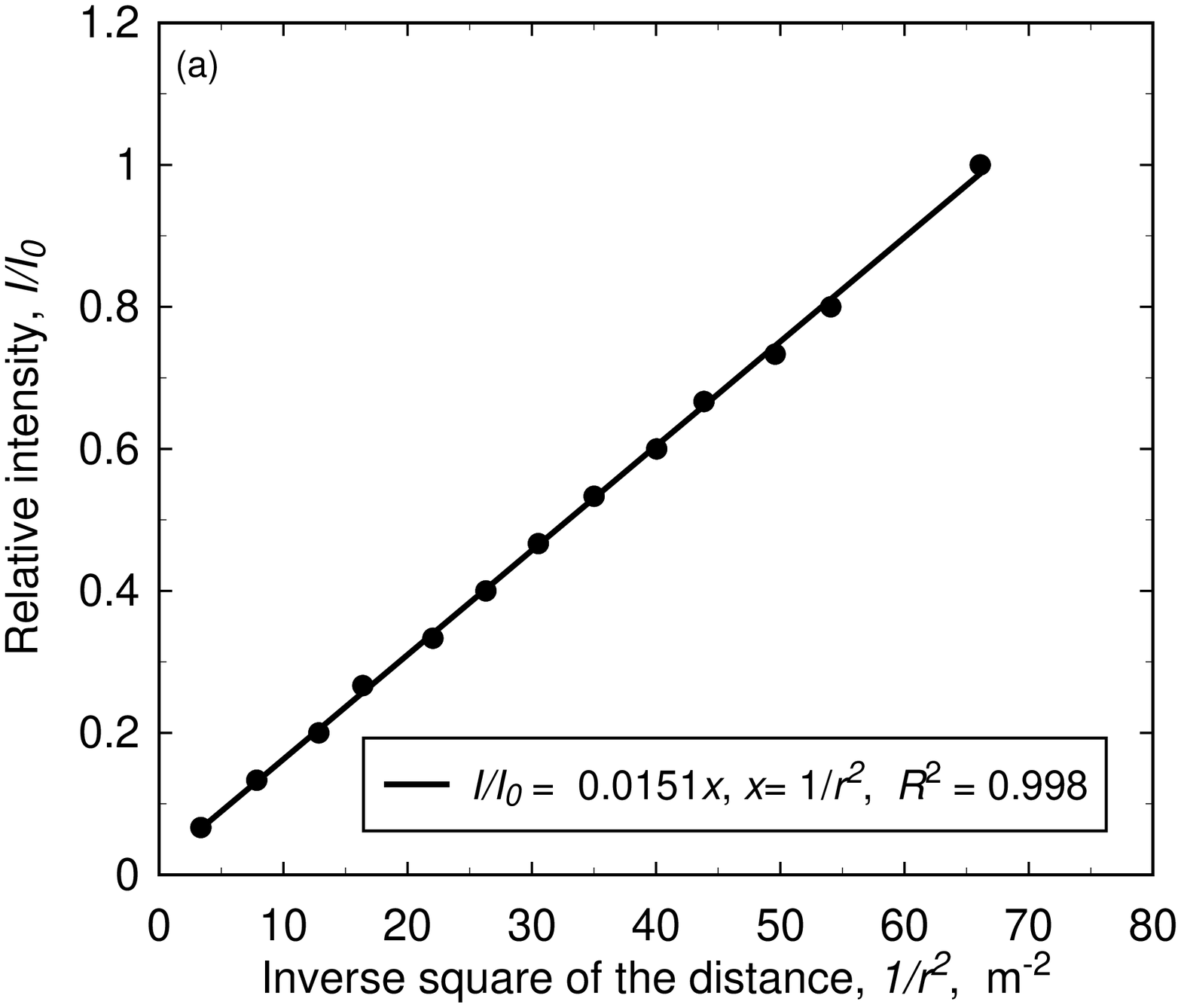}
\label{fig1a}
\vspace{-35mm}
\end{figure}
\begin{figure}[tbp]
\centering
\includegraphics[width = 85mm]{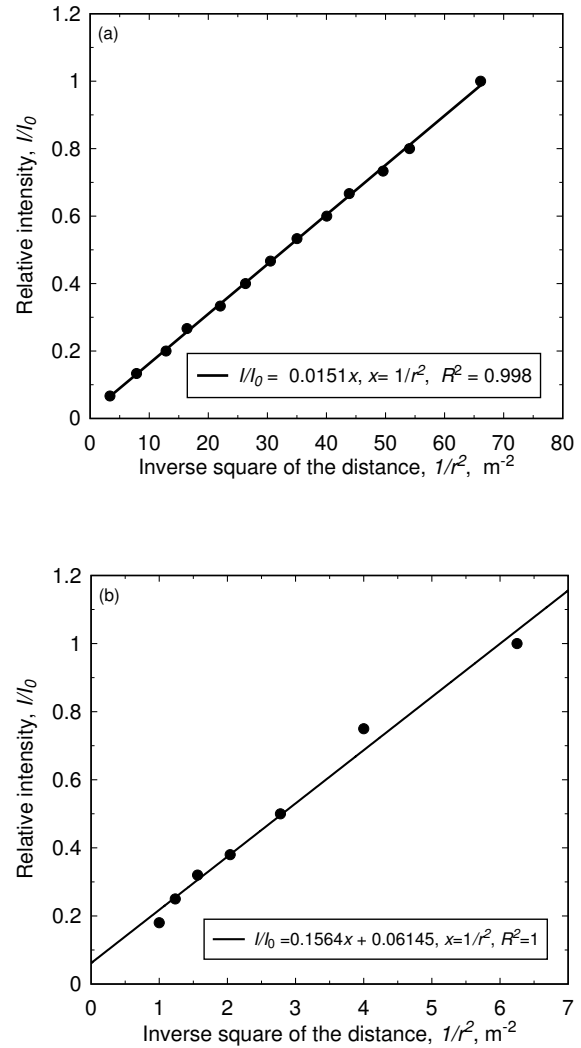}
\vspace{-25mm}
\caption{Inverse square law for light (a) and microwaves (b).}
\label{fig1b}
\end{figure}
In the case when the students perform the measurements for the
intensity of electromagnetic waves by using the microwave transmitter and
receiver they gradually increase the distance between the transmitter and
receiver by moving the receiver. Students represent the results of the
measurements for the dependence of intensity on the distance for light as
well as for microwaves in graphical form by plotting the relative intensity
versus the inverse square of the distance as shown in Figs. 1a and 1b. The
analysis of these graphs shows that intensities of visible light as well as
invisible microwaves decrease inversely proportional to the square of the
distance from the source.

\section{Reflection and Refraction of Light and Microwaves.}

\qquad There are two fundamental laws of geometric optics: the law of
reflection and Snell's law -- the law of refraction. Today's laboratory
class technology allows students to verify these laws using a beam of light
as well as a beam of monochromatic laser light. Students can perform
experiments for the reflection and refraction of light gradually changing
the angle of incidence and measuring the angle of reflection or the angle of
refraction and at the same time visually observing the propagation of the
incident ray and reflected or refracted rays for each of the incident
angles. By plotting the graph of dependence of the angle of reflection
versus the angle of incidence they can find that the slope of this graph is
unity and therefore, conclude that the angle of incidence equals the angle
of reflection. In the same way by plotting the graph of the sine of the
angle of incidence versus the sine of the angle of refraction students see
that there is a linear dependence and from the slope of the graph determine
the index of refraction for the given medium. Fig. 2a represents an example
of such dependence for refraction of light in glass. We are suggesting
studying reflection and refraction of microwaves in parallel with these
optics experiments. The difference of the setting for these experiments is
that in the case of light students actually see the reflected and the
refracted rays, but in the case of microwaves the reflected and refracted
electromagnetic waves are invisible and students determine the angle of
refraction as well as the angle of refraction of microwaves by finding the
maximum intensity for the reflected or refracted microwave radiation. Using
a transmitter and receiver of microwaves and a metallic reflected plate and
gradually increasing the angle of incidence, students can find the angle of
refraction which corresponds to the maximum intensity of reflecting
microwaves. In the case of refraction the incident microwaves are refracted
on the prism mold filled with styrene pellets. The angle of the refracted
microwaves can be found by the maximum intensity meter reading of the
refracted waves. Plotting the same graphs as in the case of light
experiments for the angle of incidence versus the angle of reflection and
for the sine of the angle of incidence versus sine of the angle of
refraction, students can verify the laws of reflection and refraction for
the microwaves and justify that these are the same as for light. 
\begin{figure}[tbp]
\centering
\includegraphics[width = 85mm]{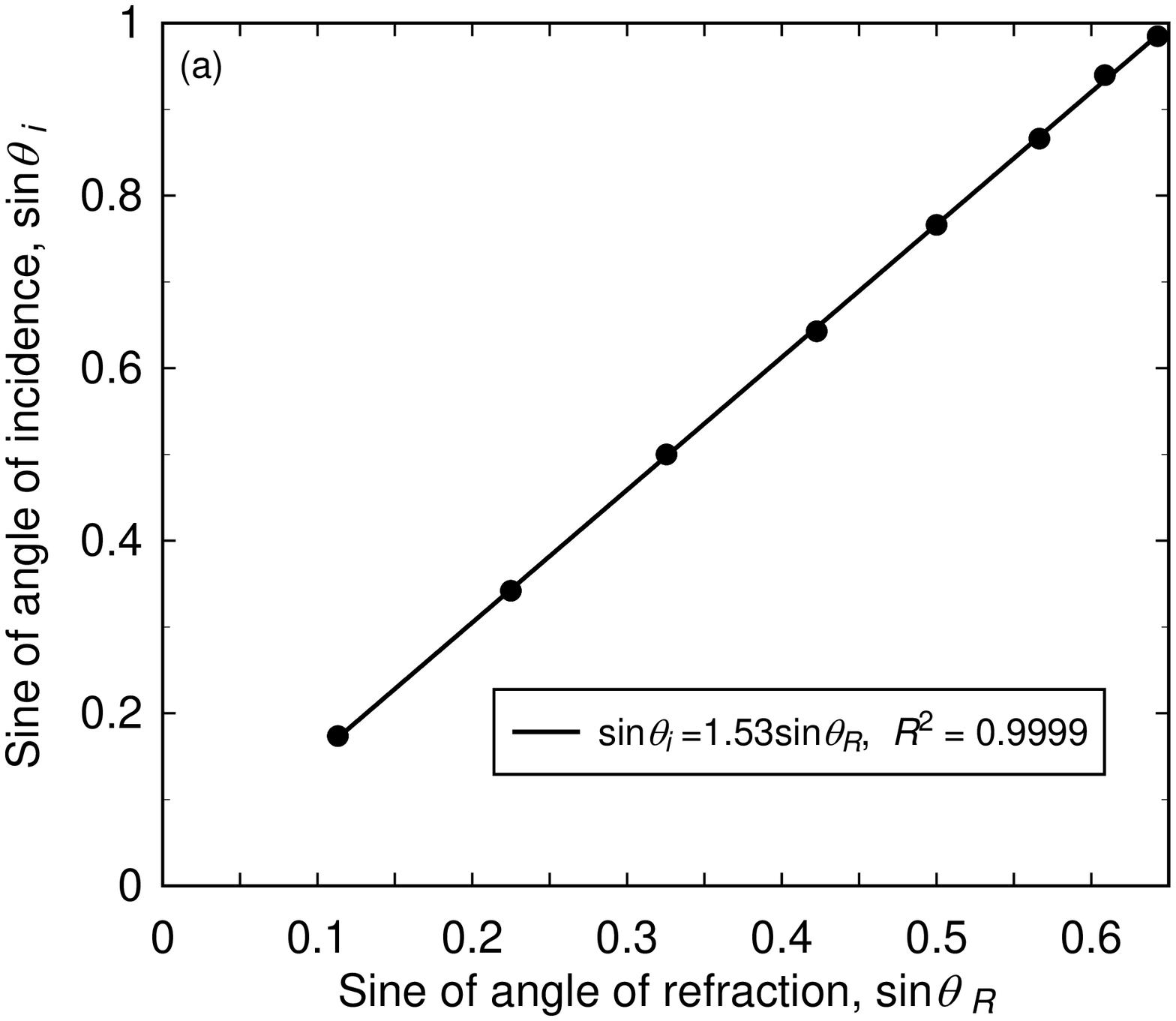}
\label{fig2a}
\vspace{-35mm}
\end{figure}
\begin{figure}[tbp]
\centering
\includegraphics[width = 85mm]{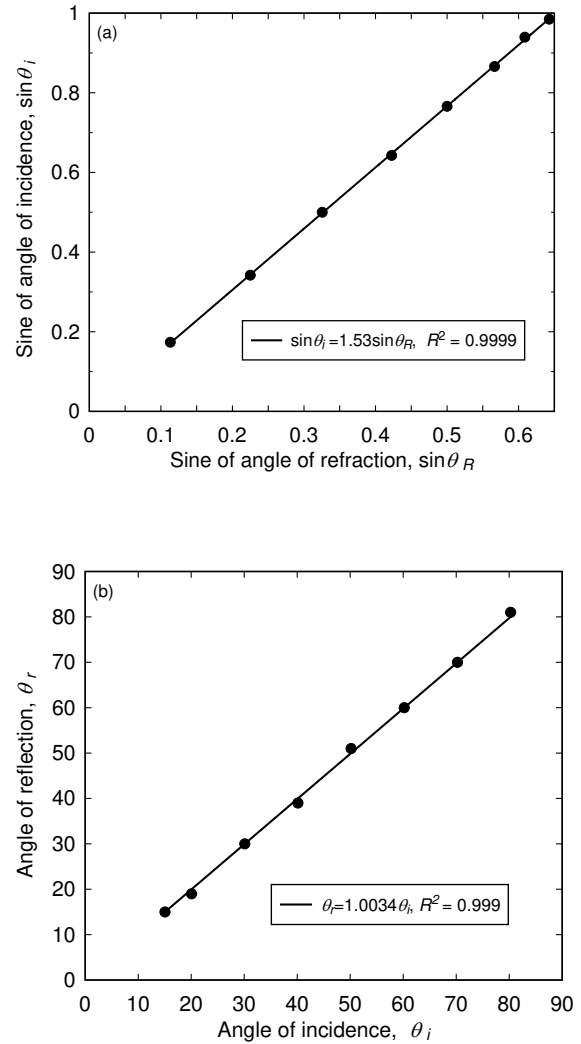}
\vspace{-25mm}
\caption{(a) Refraction of light in glass. Dependence of the sine of the angle of refraction on the sine of
the angle of incidence. (b) Reflection of microwaves. Dependence of the angle of reflection on the
angle of incidence.}
\label{fig2b}
\end{figure}
Fig. 2b presents the results for these kinds of measurements for reflection of the
microwaves.

\section{Double-Slit Interference for Light and Microwaves}

\qquad The other experiment in optics which is easy to visualize is
double-slit interference for light. This is the standard set which is
available on the market. Performing this experiment students can see a clear
interference pattern. Of course, there are many different experiments, which
also demonstrate the interference of light and visualize the interference
concept for light. We are choosing the double-slit interference of light
because a somewhat similar phenomenon occurs when microwaves pass through a
two-slit aperture and can be easily set and performed with microwaves. When
incident microwaves from a transmitter radiate on a double-slit aperture,
the intensity of the microwave beyond the aperture will vary depending on
the angle of detection by a receiver. For two thin slits separated by a
distance d, maxima of the intensity will be found at such angles that
\begin{equation}
d\sin \theta =m\lambda ,
\end{equation}%
where $\theta $\ is the angle of detection, $\lambda $\ is the wavelength of
the incident radiation, and m is any integer. Gradually changing the angle
by 5$^{0}$ for the detection position of the receiver the student measures
the intensity of microwaves beyond the double-slit aperture. Fig. 3 presents
an example of such measurements for two different runs. The solid curve
represents the measurements when the receiver is positioned close to the
double-slit aperture, while the dotted curve corresponds to the measurements
when the distance between the receiver and double-slit is increased. To
analyze the results of the experiment students for the given wavelength of
microwave calculate the angles at which they would expect the maxima and
minima to occur and compare with the locations of observed maxima and minima.
\begin{figure}[tbp]
\centering
\includegraphics[width = 85mm]{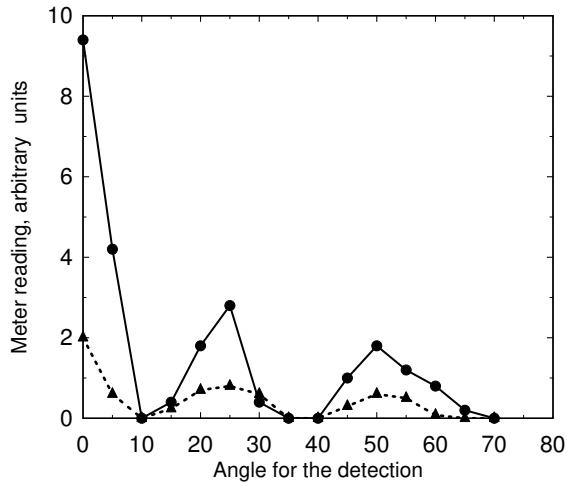}
\vspace{-25mm}
\caption{Interference pattern for microwaves.}
\label{fig3}
\end{figure}

\section{\protect\bigskip Polarization Phenomena for Light and Microwaves}

The other laboratory activities which excite students and catch their
attention deal with observation of polarization of light. Two activities
which are the easiest to set in the undergraduate college physics laboratory
are polarization by absorption and polarization by reflection. As it is well
known when unpolarized light is incident on a polarizing material, the
transmitted light is linearly polarized in the direction parallel to the
transmission axis of the polarizer. When two polarizing materials are placed
in succession in a beam of light, the first is called the polarizer and the
second - the analyzer, the amount of the light transmitted by the analyzer
depends on the angle $\theta $ between its transmission axis and the
direction of the axis of the polarizer. The intensity of light transmitted
by both polarizer and analyzer will be given by Malus's law

\begin{equation}
I=I_{0}\cos ^{2}\theta ,
\end{equation}

where $I_{0}$ is the intensity of the light incident on the analyzer. The
standard setting for this experiment requires the light source (regular
light source for qualitative observation or laser for a quantitative
observation), polarizer and analyzer to be placed in succession in a beam of
light and screen or photometer for measurement of the intensity of
transmitted light. Students set up a polarizer-analyzer system and the laser
light source as shown in Fig. 4 and orient and align their polarization
axes. Slowly rotate the analyzer in either direction while observing the
intensities of the light spot on the screen. Their observation shows that
intensity of light changes. Afterwards, students attach the analyzer to the
special component carrier of the angular translator connected to the
photometer with the fiber probe and place it instead of the screen. A
student starts at $\theta =0^{0}$ and rotates the analyzer by increasing the
angle in 10$^{0}$ increments up to 180$^{0}$ and measures the intensity of
transmitted light for the different angles of polarization. The typical
results of the measurements are presented in graphical form as shown in
Fig.5a. Then students study the polarization phenomenon for microwave
radiation. The microwave radiation from the transmitter is already linearly
polarized along the transmitter diode axis. Therefore, as the microwave
propagate through space, its electric field remains aligned with the axis of
the diode. Thus, if the transmitter diode was aligned vertically, the
electric field of the transmitted wave would be vertically polarized. If the
detector diode of the receiver is at an angle $\theta $ to the transmitter
diode, the receiver would only detect the component of the incident electric
field that was aligned along its axis. Students rotate the initially aligned
receiver from 0$^{0}$ to 180$^{0}$ in increments of ten degrees and measure
the intensities of the microwave. Plotting the graph for the relative
intensities versus the angle students can observe the similarity with the
polarization of light. By comparing the data in Figs 5a and 5b and plotting
the graph of the relative intensity versus as shown in Fig. 6 student can
conclude that the intensity of polarized light as well as polarized
microwaves follows Malus's law.
\begin{figure}[tbp]
\centering
\includegraphics[width = 85mm]{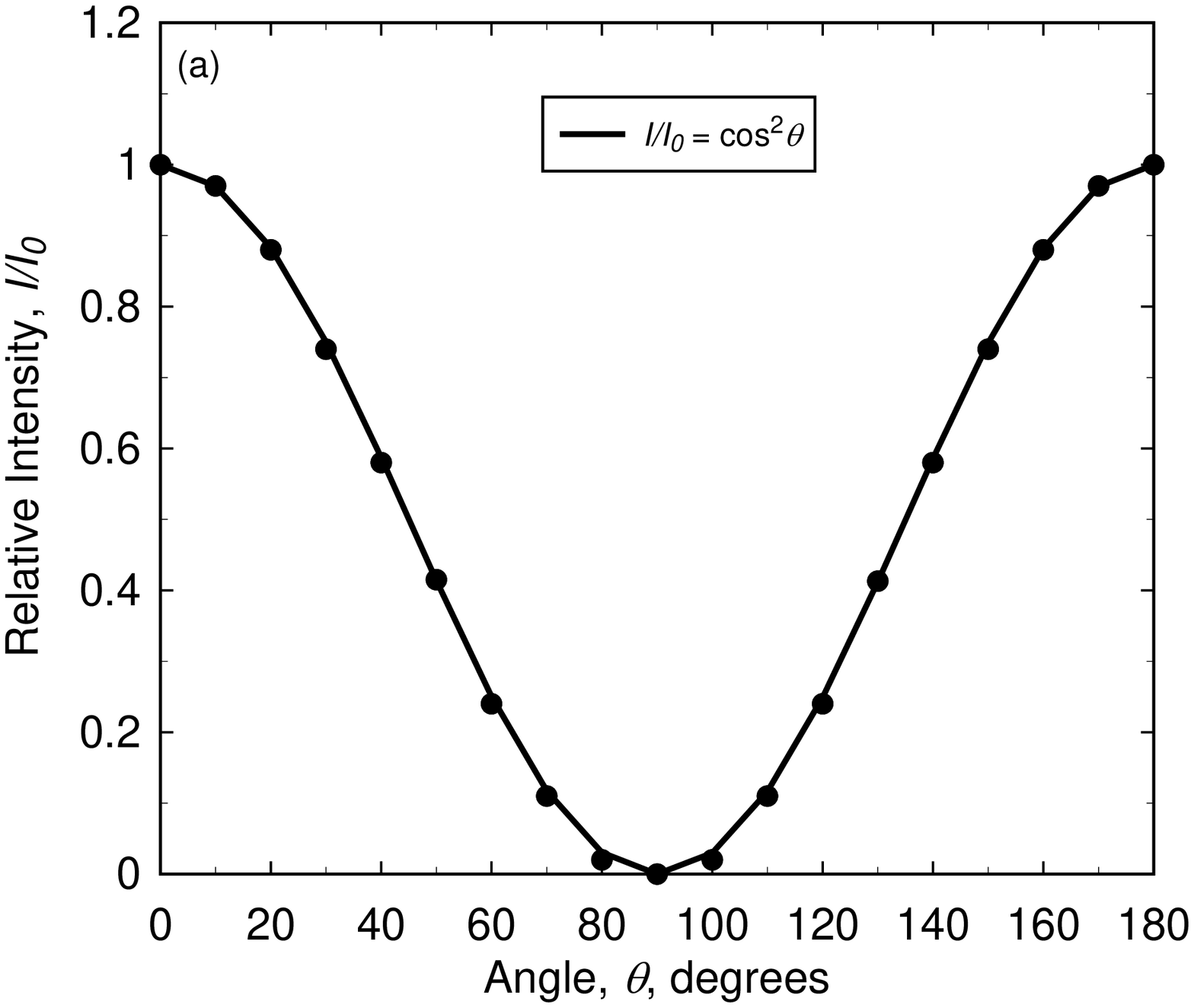}
\label{fig4a}
\vspace{-35mm}
\end{figure}
\begin{figure}[tbp]
\centering
\includegraphics[width = 85mm]{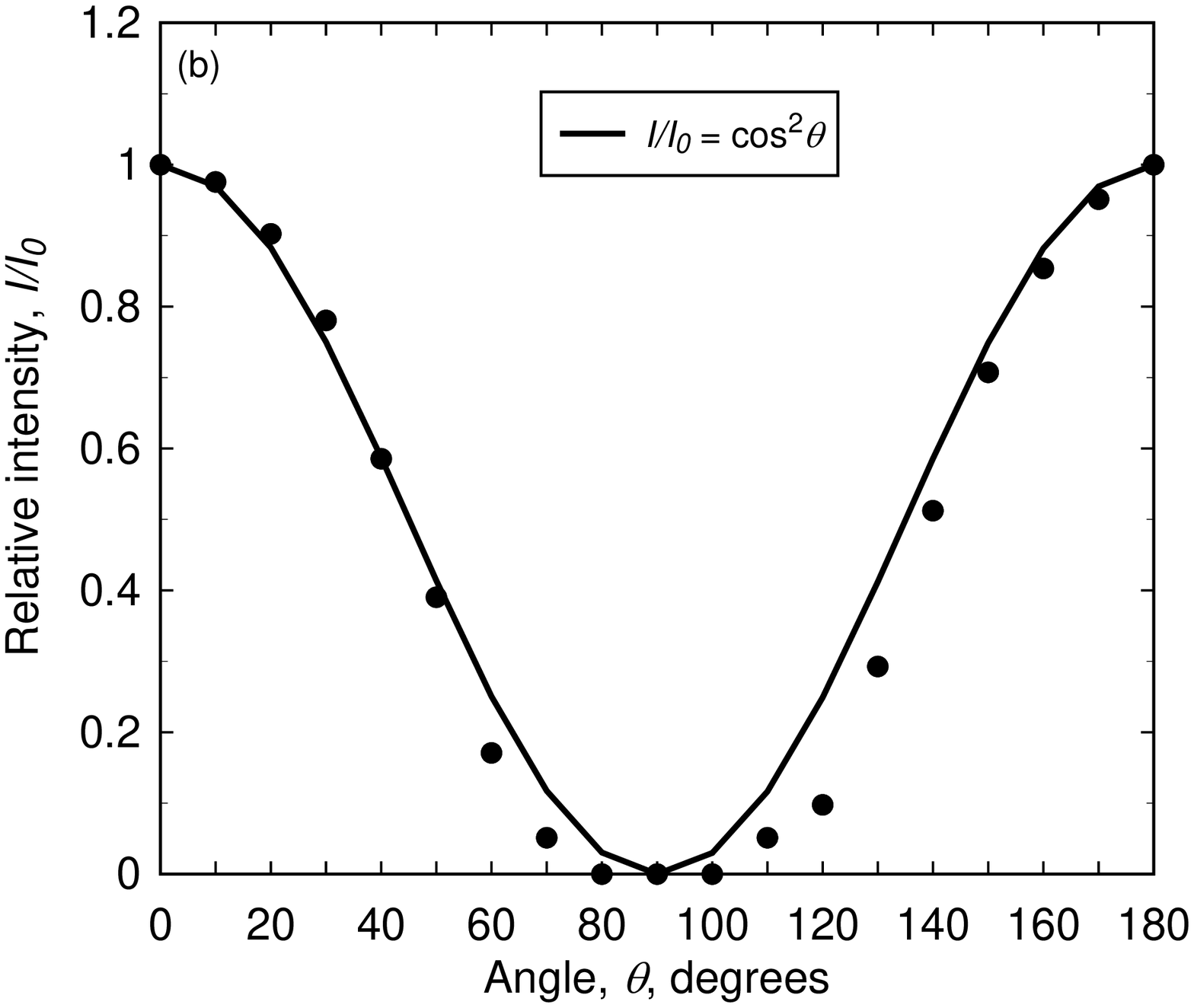}
\vspace{-25mm}
\caption{Dependence of the relative intensity of the polarized light (a) and microwaves (b) on the polarization angle.}
\label{fig4b}
\end{figure}

To better understand the mechanism of polarization for the experiment with
microwaves, student places a grid of parallel conducting strips under
different angles between the transmitter and aligned receiver and again
measures the intensities for the different angular positions of the grid.
Linearly polarized microwaves are sent through a grid of parallel conducting
strips. We choose the two perpendicular directions used to represent the
linearly polarized incident beam to be parallel and perpendicular to the
metallic strips. The polarized waves with electric field vector parallel to
the conducting strips are absorbed by the strips. The oscillatory field
parallel to the strips transfers energy to the electrons that can move along
the strips; it is the polarization direction perpendicular to the strips
that is transmitted. So the metallic strip grid acts as a polarizer, a
device for producing polarized microwaves. The axis of a polarizer is the
direction parallel and antiparallel to the plane of polarization of the
transmitted waves. The axis of a polarizer is not a unique line but simply a
direction or a whole collection of lines oriented parallel to each other.
Therefore, for the metallic strip grid polarizer of microwaves, the axis of
the polarizer is a direction in the plane of the polarizer perpendicular to
the direction of the strips. Thus, the experiments of polarization of light
and microwaves are complementary which helps to understand and visualize the
phenomenon and mechanism of polarization.
\begin{figure}[tbp]
\centering
\includegraphics[width = 85mm]{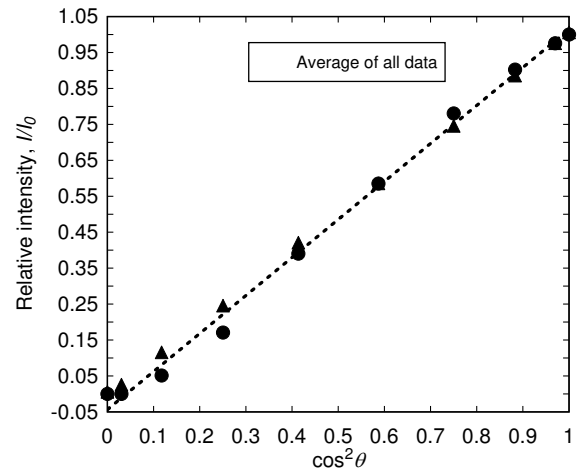}
\vspace{-25mm}
\caption{Relative intensity for polarized light and microwaves versus $\cos^{2}\theta$. Triangles represent experimental data for light and circles --- for microwaves.}
\label{fig5}
\end{figure}

Hence, on one hand students visualize the phenomenon of polarization by
observing the polarization of light. They can observe that by changing the
angle between the polarizer and analyzer the brightness of the spot of light
changes from maximum bright to dark when the angle is changing from 0$^{0}$
to 90$^{0}$. On the other hand, polarization of microwaves helps to
visualize the mechanism of polarization by observing that intensity changes
depending on the angular position of the grid.

\section{Interferometer Measurements for Wavelength of Light and Microwaves}

\qquad In the interferometry technique we superpose (interfere) two or more
electromagnetic waves, which creates an output wave different from the input
waves. Because interference is a very general phenomenon with waves,
interferometry can be applied to a wide variety of electromagnetic waves,
including optical spectrum and microwave, and can be used for measurements
of wavelength of light as well as microwaves. The Michelson interferometer
is the most common configuration for optical interferometry and can be very
easily adapted for microwave interferometry. In many scientific and
industrial uses of interferometers, a light source of a known wavelength is
used to measure incredibly small displacements - about 10 -6 meters.
However, if you know the distance of mirror movement, you can use the
interferometer to measure the wavelength of a light source as well as a
source of other electromagnetic waves. The aim of this experiment is to make
the students familiar with the simplest type of interferometers and use the
interferometer to measure the wavelength of a helium-neon laser light source
and a microwave source. The first part of the experiment is centered at
giving the students a "feeling" for the sensitivity of a Michelson
interferometer and the different types of interference patterns which can be
observed visually with the laser source. The Michelson interferometer
produces interference fringes by splitting a beam of monochromatic laser
light by a partially transparent reflector so that one beam strikes a fixed
mirror and the other a movable mirror. When the reflected beams are brought
back together, an interference pattern results. By measuring the distance $%
d_{m}$, that the movable mirror moved toward the beam-splitter and the
corresponding number of fringes $m$, students are able to determine with
high precision the wavelength of the laser light as

\begin{equation}
\lambda =\frac{2d_{m}}{m}
\end{equation}

The advantage of this part of the experiment is that the students visually
see the interference pattern. This presents a way of direct understanding of
important concepts in wave optics. In the second part of this experiment
students use the Michelson interferometer that is setup with a microwave
transmitter, partial reflector, two metallic reflectors and receiver.
Microwaves travel from the transmitter to the receiver over two different
paths. In one path, the microwave passes directly through the partial
reflector, reflects back to the first reflector, and then is reflected from
the partial reflector into the receiver. In the other path, the microwave
reflects from the partial reflector into the second reflector, and then back
through the partial reflector into the receiver. If in the optical part of
this experiment student can visually observe these pathways, for microwaves
the paths are invisible. Now by moving one of the reflectors the student
changes the path length of one wave, thereby changing its phase at the
receiver. While watching the meter, and slowly moving the reflector,
students can observe relative maxima and minima of the meter deflections. By
measuring the reflector's displacement distances and corresponding numbers
of maximum relative intensity the wavelength of microwave radiation can be
determined using the same Eq. (3) as for visible laser light.

\qquad The other advantage of this experiment is that students learn and
understand why Michelson's interferometer has become a widely used
instrument for measuring extremely small distances by using the wavelength
of a known light source. Based on this experiment students understand that
resolution for measurement of distances depends on the wavelength of
electromagnetic radiation and why an optical interferometer (an
interferometer using visible light rather than microwaves) provides better
resolution when measuring distance than a microwave interferometer.

\qquad The detailed procedures of some experiments discussed in this article
are presented in Refs.[11] and [12]. Some of the experiments are
computer-based and some are performed in the traditional way. Because not
every lab has computer access, computer-assisted experiments serve as
supplements to the traditional experiments. This allows instructors to find
the appropriate balance between traditional and computer-based experiments
for teaching topics of light and electromagnetic waves in the physics
laboratory.

\section{Conclusion}

Our ultimate goal is to improve engineering major students' learning and
motivation in physics courses by presenting the abstract material in the
traditional curriculum through real-time experiments in a laboratory
session. This approach assumes the experimental study of the properties of
light and microwaves in parallel. One of the important educational
advantages of the simultaneous study of electromagnetic waves and light is
to show that light and electromagnetic radiation have the same properties so
that the students can visualize the properties of the electromagnetic
radiation through observation of light propagation. In other words, the
abstract invisible properties of microwaves are visualized via observing
visible properties of light. On the other hand, the observation of the
polarization of microwaves helps to visualize and understand the mechanism
of the polarization of light. It is important to underline that all this is
real-time visualization but not computer-based animations, simulations or
interactive multimedia design cases. In our approach we suggested studying
the properties of electromagnetic microwave radiation and light in parallel
by performing the same laboratory experiments for light and microwaves to
show that these two phenomena demonstrate the same properties and follow the
same laws. By performing these experiments students become active
participants of the learning process.

\qquad The other advantage in teaching light and microwaves in parallel is
that students realize that for measuring the same properties of
electromagnetic waves of different range of the electromagnetic spectrum you
should use different equipment and different precision of measurements. In
our approach by analyzing data for the same phenomenon through two different
methods of measurements, students gain a greater understanding of the
concepts behind the experiments.

\qquad The visualization of complex phenomena does not by itself reach our
goal. We put this visualization into a context of the same invisible
phenomena and created a series of parallel activities that involve questions
to motivate the students and investigations of practical devices.

\acknowledgments
This work is supported by US Department of Education under the Grant P120A060052.


\begin{thebibliography}{99}
\bibitem{1} R.K. Thornton, Changing the Physics Teaching Laboratory: Using
technology and new approaches to learning to create an experimental
environment for learning physics concepts, Proceedings of the Europhysics
Conference on the Role of Experiment in Physics Education, 1997; L.C.
Epstein, Thinking physics. 1987, San Francisco: Insight Press.

\bibitem{2} G. Ippolitova, Physics and Professional Teaching in Technical
Universities, Proceedings of the International Conference on Engineering
Education August 6 -- 10, 2001 Oslo, Norway 6E5-12- 6E5-14.

\bibitem{3} J. Bernhard, Improving engineering physics teaching - learning
from physics education research. Proceedings of PTEE2000 "Physics Teaching
in Engineering Education", Budapest 13 - 17 June 2000.

\bibitem{4} R.K. Thornton, Tools for scientific thinking -
microcomputer-based laboratories for teaching physics. Phys Ed, 1987. 22: p.
230-238.

\bibitem{5} R.K. Thornton, Tools for scientific thinking, Learning physical
concepts with real-time laboratory measurements tools, in Proc Conf
Computers in Phys Instruction, E.F. Redish, Editor. Addison Wesley: Reading.
p. 177-189, 1989.

\bibitem{6} D.R. Sokoloff, R.K. Thornton, and P.W. Laws, Real-Time Physics,
active learning laboratories. 1998, New York: Wiley.

\bibitem{7} R.R. Hake, Socratic Pedagogy in the Introductory Physics Lab.
The Physics Teacher, 1992. 30: p. 546.

\bibitem{8} J.C. Maxwell, Treatise on Electricity and Magnetism. V. 2,
Oxford, 1973.

\bibitem{9} PASCO Catalog, Physics -- 2007, Science and Engineering
Education. p. 285.

\bibitem{10} DAEDALON Experiments and Apparatus Catalog 2001. p. 66-66.

\bibitem{11} R. Ya. Kezerashvili, \textquotedblleft College Physics
Laboratory Experiments. Electricity, Magnetism, Optics\textquotedblright ,
Gurami Publishing, New York, 2003.

\bibitem{12} J.D. Wilson, C. Hernandez Hall, \textquotedblleft Physics
Laboratory Experiments\textquotedblright , 6th Edition, Houghton Mifflin,
2005.
\end{thebibliography}
\end{document}